\documentclass[prb,aps,superscriptaddress,twocolumn,showkeys,floatfix]{revtex4-1}

\usepackage{times}
\usepackage{graphicx}
\begin{document}
\title{Magnetism in nanometer-thick magnetite}

\author{Matteo Monti}
\affiliation{Instituto de Qu\'{\i}mica-F\'{\i}sica ``Rocasolano'', CSIC, Madrid 28006, Spain}
\author{Benito Santos}
\affiliation{Instituto de Qu\'{\i}mica-F\'{\i}sica ``Rocasolano'', CSIC, Madrid 28006, Spain}
\author{Arantzazu Mascaraque}
\affiliation{Universidad Complutense de Madrid, Madrid 28040, Spain}
\author{Oscar Rodr\'{\i}guez de la Fuente}
\affiliation{Universidad Complutense de Madrid, Madrid 28040, Spain}
\author{Miguel Angel Ni\~{n}o}
\affiliation{Elettra Sincrotrone S.C.p.A, Trieste, Italy}
\altaffiliation{Present address: Instituto Madrile\~{n}o de Estudios Avanzados en Nanociencia (IMDEA), Madrid, Spain}
\author{Tevik Onur Mente\c{s}}
\affiliation{Elettra Sincrotrone S.C.p.A, Trieste, Italy} 
\author{Andrea Locatelli} 
\affiliation{Elettra Sincrotrone S.C.p.A, Trieste, Italy}
\author{Kevin F. McCarty}
\affiliation{Sandia National Laboratories, Livermore, CA 94550, USA}
\author{Jos\'{e} F. Marco}
\affiliation{Instituto de Qu\'{\i}mica-F\'{\i}sica ``Rocasolano'', CSIC, Madrid 28006, Spain}
\author{Juan de la Figuera}
\affiliation{Instituto de Qu\'{\i}mica-F\'{\i}sica ``Rocasolano'', CSIC, Madrid 28006, Spain}
\email{juan.delafiguera@iqfr.csic.es}

\begin{abstract}
    The oldest known magnetic material, magnetite, is of current interest for use in spintronics as a thin film. An open question is how thin can magnetite films be and still retain the robust ferrimagnetism required for many applications. We have grown one-nanometer-thick magnetite crystals and characterized them {\it in situ} by electron and photoelectron microscopies including selected-area x-ray circular dichroism. Well-defined magnetic patterns are observed in individual nano-crystals up to at least 520~K, establishing the retention of ferrimagnetism in magnetite two-unit-cells thick.

\end{abstract}

\maketitle

The trend in both magnetic data storage and spintronics is to reduce the thickness and/or lateral size of the device materials to the nanoscale. Size reduction in magnetic storage has the obvious advantage of increasing the bit density. Advantages also exist for spintronic applications. For example, the magnetic layers of spin filters can be switched with smaller magnetic fields as the layers become thinner. But size reduction can also change a material's magnetic behavior. For example, as a ferromagnet is decreased in size, at some point thermal excitations can overcome the magnetic anisotropy energy, leading to superparamagnetism. Then the material is only useful below a "blocking" temperature, where the magnetization is stable over some relevant timescale. Several approaches are used to delay the onset of superparamagnetism, including using high magnetic anisotropy materials\cite{coffey_high_1995} or by exchange bias\cite{nogues_exchange_1999}. Understanding how to stabilize magnetic order in low-dimensional structures is an important concern. 

Iron-containing oxides are a class of magnetic materials that provide good chemical stability in oxidizing atmospheres and have extremely high Curie temperatures. Strong magnetism in the iron oxide magnetite has been known since the ancient Greeks\cite{mills_lodestone_2004}. The material, historically referred to as lodestone, is currently a promising candidate for spintronic applications\cite{bibes_oxide_2007}. Bulk magnetite (Fe$_3$O$_4$) is a ferrimagnet with a 850~K Curie temperature\cite{CornellBook} and becomes multiferroic at low temperatures\cite{kato_observation_1982,alexe_ferroelectric_2009}, allowing electrical control of magnetic domains. The prediction of half-metal character\cite{katsnelson_half-metallic_2008}, which implies that the conduction electrons are 100\% spin-polarized, lead to its use as a spin injector\cite{wada_efficient_2010}.

Oxides like magnetite have much more complicated structures and larger unit cells than metal ferromagnets, giving the possibility of tuning their properties to a larger extent, specially given the often observed strong coupling to strain effects\cite{CaoMatSciEngR2011,FriakNJP2007}. Magnetite ultrathin nanostructures have been grown on a variety of substrates, including oxides\cite{IsozumiJAP1989,voogt_composition_1997,ruby_preparation_1999,chambers_epitaxial_2000,ramos_magnetotransport_2006, maris_one-dimensional_2006,parames_magnetoresistance_2007}, semiconductors\cite{reisinger_epitaxy_2003,vanderLaanPRB2004,parames_magnetoresistance_2007, boothman_structural_2007,vanderLaanPRB2010} and metals\cite{ThorntonPRB1998,weiss_surface_2002,ketteler_heteroepitaxial_2003,waddill_epitaxial_2005,santos_structure_2009,khan_nucleation_2008}. Here we examine the presence of stable ferrimagnetic domains in ultrathin magnetite, a subject with conflicting reports in the literature. While some reports state than magnetite films close to 3~nm thick present a well defined magnetic structure \cite{ThorntonPRB1998,vanderLaanPRB2004,AroraPRB2008,FernandezPachecoPRB2008,vanderLaanPRB2010}, others indicate that clear signs of superparamagnetic behavior are observed at the same thickness\cite{VoogtPRB1998,IsozumiJAP1989,KoreckiThinSol2002,HibmaJMMM2003,MoussyPRB2004}. In particular we note that, to the best of our knowledge, there are no reports of stable magnetization domains on nanometer-thick magnetite. We grow one-nanometer-thick, micron-wide, magnetite crystals on ruthenium\cite{ketteler_heteroepitaxial_2003} using reactive molecular beam epitaxy (i.e., depositing iron in a background of oxygen) while monitoring the growth in real time by low-energy electron microscopy (LEEM\cite{santos_structure_2009}). We shown by imaging their individual magnetization patterns and their x-ray magnetic circular dichroism (XMCD) spectra that they are ferrimagnetic. The observation of magnetism near the limit of unit-cell thickness shows that, under appropriate conditions, there is still room to decrease the thickness of magnetite nanostructures without introducing superparamagnetism.

\begin{figure}
\centerline{\includegraphics[width=0.45\textwidth]{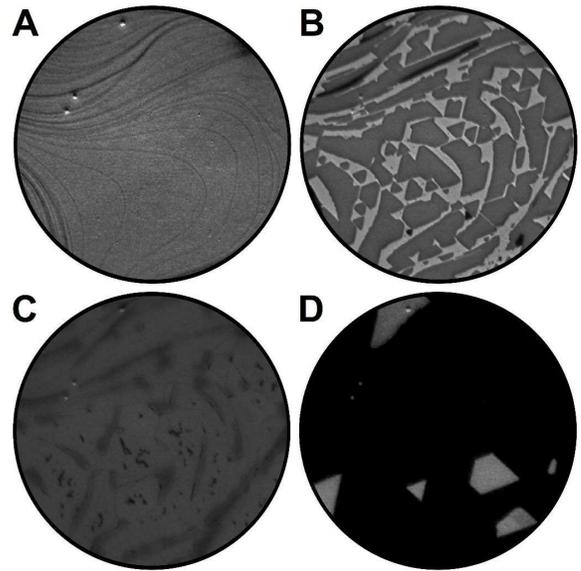}}
\caption{{(\bf A--D)} Selected LEEM images from a sequence acquired during the growth of the magnetite crystals. The first three frames show the completion of the FeO layer, while the last frame shows the final film with magnetite crystals with well-defined edges. The field of view is 10~$\mu$m and the electron beam energy is 19 eV. }
\label{growth}
\end{figure} 

The experiments were performed at the Nanospectroscopy beamline of the Elettra storage ring\cite{locatelli_photoemission_2006}. The beamline facilities include an Elmitec III  low-energy electron microscope with an hemispherical energy analyzer. The microscope has the option of selecting either an electron beam or an x-ray beam to probe the specimen surface. The electron beam allows for regular LEEM use, including fast real-space imaging of the surface during growth of the oxide films and selected-area diffraction measurements. In photoemission microscopy mode (PEEM), the instrument is able to record  selected-area x-ray absorption spectra using the secondary electrons emitted subsequent to the x-ray absorption process, or spatially resolved photoemission images. The capability of selecting the polarization of the x-ray beam (plus or minus circular polarization) allows XMCD measurements.The x-ray beam is fixed relative to the sample at an angle of 16$^\circ$ with the film plane, so the x-ray XMCD measurements are mostly sensitive to the in-plane magnetization.

The Ru single-crystal substrate with (0001) orientation was cleaned by exposure to 5$\times10^{-8}$ mbar of molecular oxygen at 1000~K, followed by flashing to 1500~K in vacuum. The sample was oriented so that the incoming x-ray beam was aligned along a mirror plane of the Ru surface (i.e., along a [11$\bar{2}$0] direction in real space). The iron oxide films were grown by reactive molecular beam epitaxy (MBE) in 5$\times10^{-7}$ mbar of molecular oxygen with the substrate at 900~K. Iron was evaporated from a 2-mm-diameter iron rod heated by electron bombardment inside a water-cooling jacket. Oxygen was introduced into the experimental chamber by means of a capillary that increased the gas flux at the sample position by about a factor of 2.

Iron oxide growth on metal substrates using molecular oxygen as the oxidizing agent is expected to occur in two stages\cite{weiss_surface_2002}: initially an FeO wetting layer covers the substrate, with a thickness that depends on the particular substrate. Then magnetite nucleates and grows as 3-dimensional islands on the FeO film. This growth mode makes it difficult to obtain ultra-thin magnetite crystals without actually imaging the film growth, as we do here. In Figure~1 several frames are presented from a sequence of LEEM images acquired during iron oxide growth. In the experiment shown, the FeO initially grows as islands comprised of two Fe-O layers. (Along its [111] direction, FeO is composed of alternating planes of iron and oxygen). When the FeO film is close to completely covering the substrate, some regions have a single FeO layer.

When further iron is deposited on a complete FeO film, large (up to several micrometer) triangular islands nucleate on top of the film (Figures~1D and 2A)\cite{santos_structure_2009}. These crystals and the FeO wetting layer are found to exhibit different low-energy electron diffraction (LEED) patterns (Fig.~2A), x-ray photoelectron spectroscopy (XPS) spectra (Fig.~2B) and x-ray absorption spectra (not shown). 

The LEED patterns of the different oxide phases are known to differ\cite{weiss_surface_2002}. The diffracted beams that arise from the periodicity of the hexagonal oxygen layers appear at very similar positions in iron oxides due to their similar oxygen-oxygen distances, 0.297--0.320~nm. However, the different arrangement of the iron atoms within the layers of each phase gives rise to 1$\times$1, 2$\times$2 and $\sqrt{3}\times\sqrt{3}$R30$^\circ$ LEED patterns for bulk-terminated FeO(111), magnetite(111) and hematite(0001), respectively. As seen in Figure~2A, the wetting layer has a 1$\times$1 LEED pattern (with additional spots due to a coincidence pattern with the underlying Ru substrate), suggesting a FeO(111) surface. In contrast, the large triangular crystal has a 2$\times$2 LEED pattern, which is indicative of magnetite. The island's oxygen lattice spacing obtained from LEED is the same as the wetting layer, 0.32$\pm$0.04~nm, i.e., the magnetite crystals are strained by 6\%.

\begin{figure}
\centerline{\includegraphics[width=0.5\textwidth]{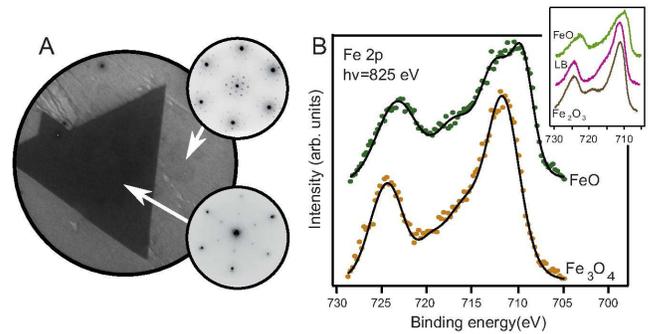}}
\caption{(color online)({\bf A}) LEEM image of a magnetite crystal (the field of view is 4 $\mu$m and the electron energy is 8~eV), with insets showing the low-energy electron diffraction patterns (acquired using an electron beam of 28~eV) of the crystal and its surrounding wetting layer. ({\bf B}) Fe $2p$ core-level x-ray photoelectron spectra acquired from the crystal (orange, lower curve) and  the wetting layer (green, upper curve). The solid lines are the sum of the different individual contributions (not shown) that are expected to be present in the XPS spectra of FeO and magnetite. The inset shows reference spectra for FeO (top, green), a Langmuir-Blodgett film containing Fe$^{2+}$ and Fe$^{3+}$ (middle, purple), and hematite (bottom, gray). A non-linear background has been substracted from the spectra.}
\label{crystals}
\end{figure}

Aimed at identifying more precisely the chemical nature of the triangular crystals, the Fe 2$p$ core level XPS spectra were recorded from a crystal and from its surrounding iron oxide layer (Figure~2B). For comparison, the inset shows the same Fe 2$p$ core level peaks recorded from several iron compounds obtained using a conventional laboratory XPS spectrometer. (The upper spectrum corresponds to an FeO film produced by vacuum evaporation of Fe metal on a Ru substrate and subsequent oxidation; the middle spectrum corresponds to a 3-nm-thick Fe-containing film produced by the Langmuir-Blodgett (LB) technique that contains both Fe$^{2+}$ and Fe$^{3+}$\cite{mercedes};  
 the bottom spectrum was obtained from pure $\alpha$-Fe$_2$O$_3$ powder.) The spectrum from the wetting layer (Figure 2B, top) presents the same features as the reference FeO film, confirming that the wetting layer is FeO. In contrast, the main photoemission peaks in the Fe 2$p$ spectrum from the triangular crystal (Figure 2B, bottom) appear at higher binding energies than those in the wetting-layer spectrum, indicating that the average Fe oxidation state in the crystal is higher than in the wetting layer. The characteristic shake-up satellite of exclusively Fe$^{3+}$-containing phases (i.e., the peak at 718--719~eV) is not evident in the crystal's spectrum. The spectrum thus resembles that of the mixed Fe$^{2+}$-Fe$^{3+}$ LB film shown in the middle of the inset of Figure~2B, indicating that the crystal is a mixed-valence Fe$^{2+}$/Fe$^{3+}$ oxide. Consistently the octahedral positions in magnetite's inverse spinel structure\cite{CornellBook} are populated with both Fe$^{2+}$ and Fe$^{3+}$ while the tetrahedral positions are occupied only by Fe$^{3+}$. This result together with the LEED pattern [and the x-ray circular magnetic dichroism (XMCD) spectra of Fig.~3D] indicate that the ultrathin crystals are indeed magnetite. We cannot, however, evaluate their detailed stoichiometry. (Magnetite is often non-stoichiometric.)

\begin{figure}
\centerline{\includegraphics[width=0.4\textwidth]{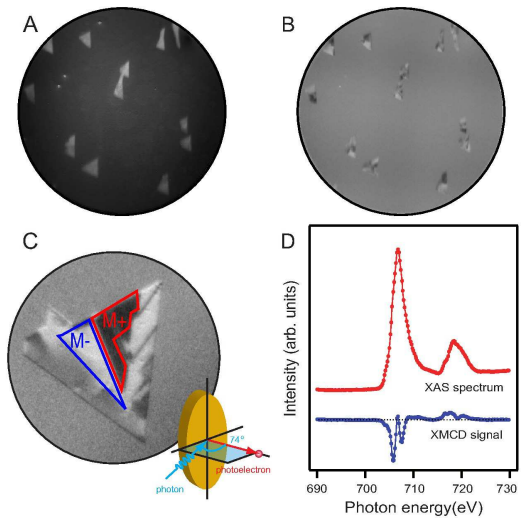}}
\caption{(color online)({\bf A}) XAS and ({\bf B}) XMCD image at 705.8 eV. The field of view is 30~$\mu$m. ({\bf C}) XMCD image recorded in remanence showing the magnetization pattern of same crystal presented in A. The field of view is 4~$\mu$m. The inset showns the experimental geometry. ({\bf D}) Top: XAS spectrum from the magnetite crystal. Bottom: XMCD difference spectrum.}
\label{XAS}
\end{figure}

We use XMCD in PEEM\cite{schneider_investigating_2002} to reveal in-situ the magnetic order of the individual magnetite crystals. To measure the x-ray absorption (XAS) spectra for the magnetite crystals, an image of the secondary electron emission (which is proportional to the x-ray absorption) was collected while the photon energy was scanned over the Fe L$_{3,2}$ x-ray absorption edges in two different  scans using opposite x-ray helicities. Such a XAS image, acquired close to the Fe L$_3$ absorption edge, is shown in Figure~3A. The image intensity from the area corresponding to the magnetite crystal of Figure~2A was integrated and averaged for the two  x-rays helicities, giving the XAS spectrum shown in Figure~3D(top). The spectrum provides further support that the crystal is magnetite\cite{crocombette_x-ray-absorption_1995}, which has a significant XMCD signal\cite{kuiper_fe_1997,huang_spin_2004,goering_vanishing_2006} at the shoulder before the maximum of the L$_3$ XAS spectra. Taking images at this photon energy (705.8 eV) with different helicities and subtracting them pixel by pixel gives the XMCD images of Figure~3B (larger field of view of 30~$\mu$m, showing the well separated magnetite crystals, all of which present magnetic domains) and Figure~3C (where the same island of Figure~2A is shown). The uniform gray intensity of the FeO wetting layer indicates that it has no magnetic circular dichroic contrast. We thus do not find any ferromagnetic order such as the one observed on FeO/Fe(110)\cite{mori}, in agreement with the antiferromagnetic order expected both in bulk FeO (which is antiferromagnetic with a N\'{e}el temperature below room temperature\cite{CornellBook}) and in an ultra-thin FeO film on Pt(111)\cite{giordano_interplay_2007}. In contrast, the magnetite crystals show a clear dichroic contrast, establishing that they have non-zero local magnetization. The magnetic domain patterns (Fig.~3C) are intricate, with long straight domain walls oriented along the \{112\} directions of the magnetite crystal. The two domains marked as $M+$ and $M-$ in Figure~3C have the similar magnitude of  the magnetization component along the illumination direction of the x-ray beam. The magnetization pattern persists during annealing up to 520~K, where changes in the surface topography are already detected. 

In order to calculate the dichroic XMCD spectra, only the area that corresponds to a given domain in a XAS image with a given helicity is selected: a different XAS spectra can be collected for each specific combination of domain type ($M+$, $M-$) and x-ray polarization ($P+$, $P-$). To avoid spurious signals, the $I(+M,+P)$ and $I(-M, +P)$ curves were subtracted together, as were the $I(+M,-P)$ and the $I(-M,-P)$ curves. Then each of the two difference spectra for $+P$ and $-P$ are subtracted from one another, after normalizing by the difference in XAS intensity in the wetting layer to account for illumination differences. The XMCD difference spectrum (Figure~3D, bottom) shows a well-defined peak structure at the L$_3$ and  L$_2$ edges that is characteristic of magnetite\cite{goering_vanishing_2006}. The two negative peaks at the L$_3$ edge originate mostly from the iron cations sitting at the octahedral sites, with nominal valences +2 and +3. The positive peak in the middle corresponds mostly to the tetrahedral Fe$^{3+}$ ions. The opposite sense of the XMCD peaks from the iron cations at octahedral and tetrahedral sites indicates their mutual antiferromagnetic coupling. Thus the magnetite crystals are ferrimagnetic, like the bulk material. The tetrahedral peak is smaller and the octahedral Fe$^{2+}$ peak is larger than for bulk magnetite\cite{goering_vanishing_2006}. These small differences may arise from contributions from the underlying FeO wetting layer\cite{ingo} or from an incomplete unit cell\cite{AroraPRB2008}. 

\begin{figure}
\centerline{\includegraphics[width=0.4\textwidth]{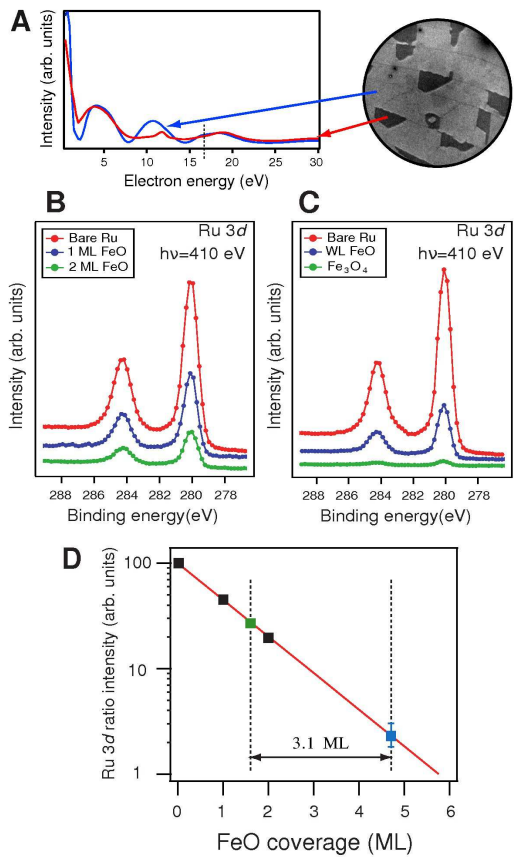}}
\caption{(color online)Determination of the magnetite crystal thickness.  (\textbf{A}) Left: Electron reflectivity curves recorded from two regions of a continuous FeO film, in blue for the majority area (which appears light gray in the LEEM image on the right side and corresponds to a FeO bilayer) and in red for the minority regions (which appear in dark gray in the LEEM image and correspond to a FeO single layer). Right: LEEM image of a complete FeO layer before growth of the magnetite crystals. (The electron energy is 16.75~eV and the field of view is 4~$\mu$m). ({\bf B}) Ru 3$d$ XPS spectra recorded from clean Ru, through a FeO monolayer and bilayer, respectively. ({\bf C}) Ru 3$d$ XPS spectra recorded from clean Ru, through the wetting layer around the magnetite crystals, and through the magnetite crystals, respectively. ({\bf D}) Semilogaritmic plot of the relative Ru XPS 3$d_{5/2}$ peak area recorded from the different films versus coverage in FeO layers.  The black squares correspond to the spectra in B, which give a mean free path of 1.25$\pm$0.03~ML$_{FeO}$ (line in the semilogaritmic plot). The green and blue squares correspond to the wetting layer and the magnetite crystal, respectively, from C (error bars from the wetting layer are within the symbol size). The additional thickness of the magnetite crystal does not depend on the particular wetting layer thickness.}
\label{attenuation}
\end{figure}

We next accurately measure the thickness of individual magnetite crystals. We use the 40-nm real-space resolution of PEEM to measure the attenuation of the photoelectrons emitted from the Ru $3d$ core level of the substrate when emerging through individual magnetite crystals (Fig.~4). This method requires an accurate value of the mean free path of electrons traveling through the magnetite crystal at a given kinetic energy. For  400~eV photons the electrons from the Ru 3$d_{5/2}$ core level have a kinetic energy of 120~eV. The attenuation of a single FeO layer was measured by comparing the spectral area of the Ru 3$d_{5/2}$ core level from bare Ru, measured through a FeO bi-layer and through a single FeO layer (see Fig.~4). The bilayer and single layer areas of FeO are easily distinguished not only by the difference in the substrate core level attenuation but by their electron reflectivity (Figure~4A). The FeO bilayer presents an additional peak absent from the FeO monolayer. (Oscillations in electron reflectivity with energy arise from interference between electrons reflected from the film/substrate and film/vacuum interfaces\cite{altman_quantum_2001}, with each additional peak indicating one additional layer.) Using the experimentally determined mean free path for 120~eV electrons in FeO (1.25$\pm$0.02 FeO layers, Figure~4D), the thickness of the magnetite crystal was estimated to be 3.1$\pm$0.3 FeO layers (see Fig.~4C). Given the relative density of the Fe-O layers in magnetite and in FeO (same oxygen density, 25\% smaller iron density for magnetite), and the thickness of the Fe-O layers in magnetite [0.242$\pm$0.06~nm\cite{weiss_surface_2002}] yields a thickness for the magnetite crystals of 1.0$\pm$0.4~nm (the larger relative error for the thickness in nm is due to the conservative estimate of the influence of the relative density of magnetite Fe-O layers in magnetite and FeO). To put this number into context we note that the thinnest magnetite film grown on a metal substrate (with a similar FeO interface layer\cite{weiss_surface_2002}) reported to date with a well defined bulk-like local magnetic structure, determined by Conversion Electron M\"ossbauer spectroscopy, was around three times thicker that our islands\cite{ThorntonPRB1998}. In contrast, thinner films often show a marked superparamagnetic behavior such as for 1.8~nm magnetite on spinel films reported by Eerenstein et al.\cite{HibmaJMMM2003}. Anti-phase boundaries (APB) are often blamed for the appearance of superparamagnetism \cite{BerkowitzPRL1997,MoussyPRB2004}. Our films seem to have a low density of APB (as detected from the size of the magnetic domains detected as well as from previous dark field imaging -see Figure~7 in Ref.\onlinecite{santos_structure_2009}-), which might account for the stable magnetization domains on our thinner magnetite crystals.  

\section*{Summary}

In summary, we have grown one-nanometer-thick crystals of iron oxide on a substrate. Electron diffraction, Fe core level photoelectron spectroscopy and x-ray absorption spectroscopy establish that the crystals are magnetite. X-ray magnetic circular dichroism reveals that individual ultrathin crystals have ferrimagnetic order up to 520~K. With thickness of only two unit cells, the crystals may well be the thinnest lodestone ever and they establish that magnetite's robust magnetism is preserved at the nanometer limit.

\section*{Acknowledgments}

This research was supported by the Spanish Ministry of Science and Innovation through Projects No.~MAT2009-14578-C03-01, MAT2009-14578-C03-02 and MAT2010-21156-C03-02, by the Office of Basic Energy Sciences, Division of Materials and Engineering Sciences, U.~S. Department of Energy under Contract No. DE-AC04-94AL85000, and by the European Union through 226716-ELISA. M. M. and B. S. thank the Spanish Ministry of Science and Innovation for supporting them through FPI fellowships.

\bibliography{FeO}

\end{document}